\documentclass[10pt,twocolumn,letterpaper]{article}

\usepackage[pagenumbers]{cvpr} 

\usepackage{graphicx}
\usepackage{amsmath}
\usepackage{amssymb}
\usepackage{booktabs}
\usepackage{engord}
\usepackage{bm}
\usepackage{url}
\usepackage{multirow}
\usepackage{xcolor}
\usepackage{soul}

\usepackage[pagebackref,breaklinks,colorlinks]{hyperref}

\usepackage[capitalize]{cleveref}
\crefname{section}{Sec.}{Secs.}
\Crefname{section}{Section}{Sections}
\Crefname{table}{Table}{Tables}
\crefname{table}{Tab.}{Tabs.}

\def\cvprPaperID{1919} 
\def\confName{CVPR}
\def\confYear{2023}

\begin{document}

\title{Backdoor Attacks Against Incremental Learners: \\An Empirical Evaluation Study}

\author{Yiqi Zhong$^{1}$, Xianming Liu$^{1}$\thanks{Corresponding author}, Deming Zhai$^{1}$, Junjun Jiang$^{1}$, Xiangyang Ji$^{2}$\\
$^{1}$Harbin Institute of Technology, $^{2}$Tsinghua University\\
{\tt\small 21s003117@stu.hit.edu.cn, \{csxm, zhaideming, jiangjunjun\}@hit.edu.cn, xyji@tsinghua.edu.cn}
}
\maketitle

\begin{abstract}
   Large amounts of incremental learning algorithms have been proposed to alleviate the catastrophic forgetting issue arises while dealing with sequential data on a time series. However, the adversarial robustness of incremental learners has not been widely verified, leaving potential security risks. Specifically, for poisoning-based backdoor attacks, we argue that the nature of streaming data in IL provides great convenience to the adversary by creating the possibility of distributed and cross-task attacks---an adversary can affect \textbf{any unknown} previous or subsequent task by data poisoning \textbf{at any time or time series} with extremely small amount of backdoor samples injected (e.g., $0.1\%$ based on our observations). To attract the attention of the research community, in this paper, we empirically reveal the high vulnerability of 11 typical incremental learners against poisoning-based backdoor attack on 3 learning scenarios, especially the cross-task generalization effect of backdoor knowledge, while the poison ratios range from $5\%$ to as low as $0.1\%$. Finally, the defense mechanism based on activation clustering is found to be effective in detecting our trigger pattern to mitigate potential security risks.
\end{abstract}

\section{Introduction}
\label{sec:intro}

\begin{figure*}[t]
    \centering
    \includegraphics[width=0.90\linewidth,trim=0 0 0 0]{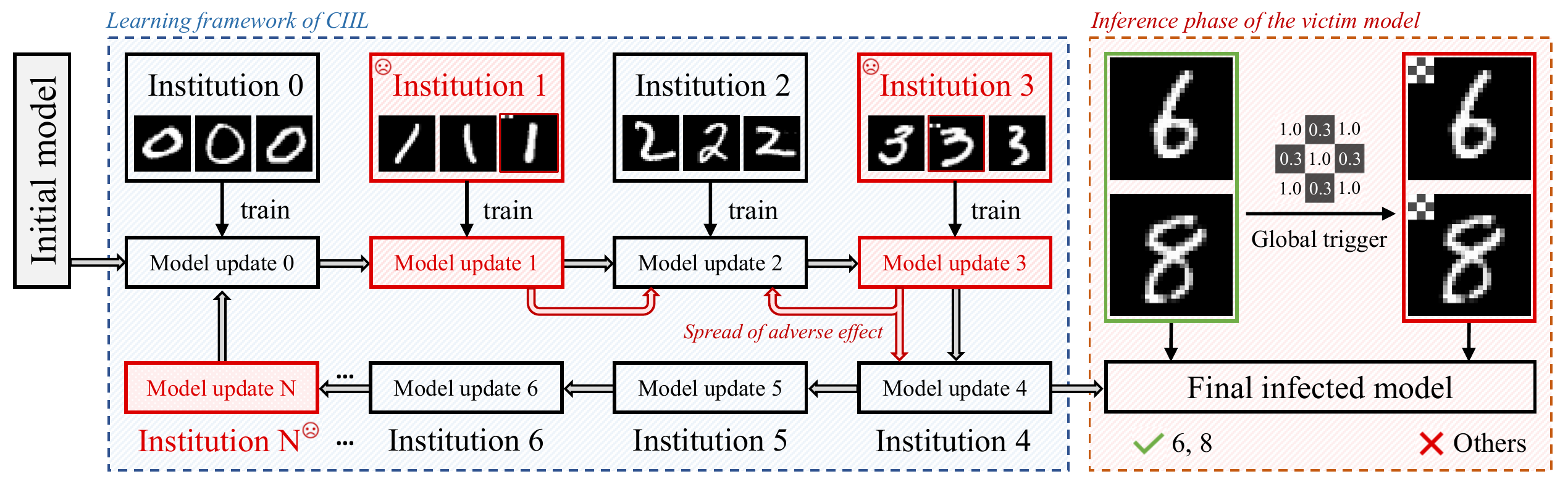}
    \caption{In the scenario of incremental learning, backdoor attacks based on data poisoning can be distributed. For instance, under the cyclic institutional incremental learning (CIIL) framework\cite{govindaswamy2021federated}, the adversary can compromise the whole system by poisoning the training set of only a few institutions due to the spread of adverse effect across institutions (tasks). We use the typical 6$\times$6 checkerboard-shaped global trigger pattern in our experiments, and follow the distributed attack described in \cite{xie2019dba}, we only keep two rows randomly from the global trigger as the local trigger for each institution (task) to further promote stealthiness.} 
    \label{fig:introduce}
\end{figure*}

In practice, we face a dynamic world, where the data and knowledge are available sequentially. The traditional neural network based learning works with static models, which are incapable of adapting their behavior over time. They have to repeat the training process each time when new data are coming. Incremental learning (IL), also referred to as lifelong learning, serves as a new learning paradigm, which allows continually accumulating knowledge without the need to retrain neural network from scratch. Due to its superiority, IL receives more attentions from diverse fields, ranging from theory study in machine learning to application study in computer vision.

In addition to handling the data stream with dynamic distribution in the real world, IL can also be use for collaborative learning among institutions without sharing private data, namely institutional incremental learning (IIL \cite{govindaswamy2021federated}). In contrast to federated learning which aggregates updates of parameters from each client by a central server, each node in IIL trains the global model incrementally and then passes the trained model to the next node. As shown in the learning framework displayed in \cref{fig:introduce}, this process can run in cycles to further alleviate the problem of catastrophic forgetting and adapt the model with newly generated training data from each node, which is called cyclic IIL (CIIL).

Up to now, most researches in IL focus on remedying the challenge of learning without catastrophic forgetting, \textit{i.e.}, the stability-plasticity dilemma \cite{de2021continual}: stability on retaining previous knowledge, while plasticity on integrating new knowledge. However, very few work considers the security issues in IL. In this paper, we present a comprehensive empirical study on three scenarios of IL to show their vulnerability to backdoor attack.

Currently, the security issues of DNNs-based learning has aroused wide concern since many real-world tasks work with high security requirements, such as autonomous driving \cite{caesar2020nuscenes} and medical diagnosis \cite{shen2017deep}. While achieving superior performance, DNNs has been shown to be brittle under well-designed adversarial attacks. There are two main lines of research in the literature to comprehensively evaluate the robustness of DNN models, namely evasion attacks during the inference phase \cite{szegedy2013intriguing} and backdoor attacks mainly before the models are finally deployed \cite{gu2019badnets}. In this work, we consider the influence of backdoor attack to IL.

To train DNN models, large-scale training datasets are necessary to achieve satisfactory performance, which however may be unattainable in many applications. People then turn to use third-party datasets or fine-tune their model upon pre-trained models directly, or even entrust the development of AI algorithms to third-party teams, such as the popular Machine Learning as a Service \cite{hunt2018chiron}. Although these strategies facilitate the training process, they introduce uncontrollable external factors, making backdoor attack possible. Backdoor attack could be conducted in any process of algorithm development, including model constructing \cite{tang2020embarrassingly}, training \cite{gu2019badnets} and deployment \cite{rakin2020tbt}. Consequently, the robustness against backdoor attack is a critical concern for machine learning system, especially for newly designed algorithms without extensive evaluation, such as IL learners.

It is crucial to comprehensively understand the vulnerabilities of incremental learners as they have been applied in safety-critical scenarios. We argue that the nature of streaming data in IL greatly increases the possibility of being attacked by a backdoor adversary. Similar to existing work on model poisoning for federated learning \cite{bhagoji2019analyzing}, in this paper, we empirically find that poisoning the training set of one task in IL (one institution for IIL) is sufficient to make the trigger work on all tasks due to the spread of adverse effect. Furthermore, as illustrated in \cref{fig:introduce}, the attack can be distributed to magnify the destructiveness. Note that there are already a few work explored the vulnerabilities of IL to data poisoning and backdoor attacks \cite{umer2020targeted,umer2022false,li2022targeted,umer2021adversarial}, which also confirmed the feasibility of cross-task attacks. However, they focus on degrading the performance of only one specific target task (\textit{e.g.}, the first task), and the poisoned samples for different target tasks may be distinctive. \textbf{In contrary, we focus on the flexibility and distributed feature, and make the trigger pattern work on all tasks rather than a specific task by utilizing a generic poisoning method.} 

Our contributions can be summarized as follows:
\begin{itemize}
    \item We comprehensively evaluated the vulnerability of incremental learning to backdoor attack on 11 incremental learners and 3 learning scenarios, demonstrating that the trigger implanted at any time can successfully work on any previous or subsequent tasks.
    \item We are the first to explore distributed backdoor attack for IL, and find that it can cause more significant performance drop with fewer poisoned samples and stealthier trigger pattern for each task, which reveals a new harmful threat.
    \item We discuss a defense strategy based on activation clustering to detect the backdoor attack, and verify its effectiveness through experiments.
\end{itemize}

\section{Related work}
\label{sec:related_work}

\subsection{Incremental learning}

Deep neural networks (DNNs) suffer from catastrophic forgetting, also known as the stability-plasticity dilemma
\cite{parisi2019continual}. In order to equip DNNs with the ability to continuously learn new knowledge while keeping their performance on existing tasks, there are generally three types of strategies in the literature, namely architecture-based, regularization-based and replay-based learners \cite{parisi2019continual,de2021continual}. 

The principle behind architecture-based learners is quit intuitive, \textit{i.e.}, the isolation of parameters can eliminate the influence between tasks, which is achieved by allocating different sub-networks in a fixed model \cite{masse2018alleviating,mallya2018packnet,fernando2017pathnet}, or by expanding the model architecture for new tasks \cite{rusu2016progressive,xu2018reinforced}. For instance, XdG \cite{masse2018alleviating} randomly assigns a subset of neurons for each task. By contrast, regularization-based learners keep old knowledge by introducing a regularization term into the loss function \cite{kirkpatrick2017overcoming,zenke2017continual,li2017learning}, among which EWC \cite{kirkpatrick2017overcoming} penalizes changes in parameters that are important to previous tasks by utilizing Fisher information matrix. Replay-based learners are jointly optimized on new samples and replayed samples which are sampled from a memory buffer \cite{rebuffi2017icarl} or a generative model \cite{shin2017continual,van2019three}. In this work, we empirically study the backdoor-related security risks of the above three types of learners.

\subsection{Backdoor attack}

In opposition to evasion attacks which focus on the inference phase, backdoor attacks are often implemented during the process of model constructing \cite{tang2020embarrassingly}, training \cite{gu2019badnets,chen2017targeted} and deployment \cite{rakin2020tbt}. 
Most of the backdoor attacks in the literature are based on data poisoning \cite{li2022backdoor, chen2017targeted, gu2019badnets, liu2020reflection, turner2019label}, among which a common practice is to inject a small number of samples with trigger patterns and possibly modified labels into the training set, namely data poisoning. 

The susceptibility of incremental learners to backdoor attacks has been investigated by a few prior work \cite{umer2020targeted,umer2022false,li2022targeted,umer2021adversarial}. Li \textit{et al}. \cite{li2022targeted} proposed a white-box task targeted data poisoning attack. They assume that the model parameters are known to the adversary, and generate $l_\infty$-norm $\epsilon$-bounded perturbations for the training set of the current task by gradient-based optimization, such that the training process on this poisoned dataset will force the model to forget the knowledge of a specified previous task. Umer \textit{et al.}\cite{umer2020targeted,umer2022false,umer2021adversarial} attacked both regularization-based and replay-based learners in Domain-IL and Task-IL scenarios with imperceptible trigger patterns, resulting in significant performance degradation of the first task. In contrast to existing work, we consider a more practical black-box setting where details of the models are unknown. We focus on the flexibility and distributed feature of attacks by thorough experiments on 11 learners and 3 learning scenarios, demonstrating that an adversary can destroy the performance of any previous or subsequent tasks by data poisoning at any time or time series, rather than targeting at a specified task.

\section{Backdoor attacks on incremental learners}
\label{sec:evaluation}

In this section, we evaluate the adversarial robustness of incremental learners, which has rarely been considered in previous works. 
Specifically, given a sequence of tasks in a time series and an incremental learner, the adversary gains an additional degree of freedom in the dimension of time compared to conventional adversarial learning scenarios. We investigate whether the implanted backdoor can be successfully triggered on any previous or subsequent tasks once we implement data poisoning at a certain time or sub-time series, which can greatly increase the flexibility and feasibility of attacks. Furthermore, it is worth noting that some strategies adopted in incremental learners may make this goal come true, such as replay strategy, since joint training of old and new samples will facilitate the propagation of backdoor knowledge across tasks. Next, we will carry out comprehensive experimental evaluation on the cross-task and distributed feature of backdoor attack in IL.

\subsection{Threat model \& Experimental setting}
\label{sec:setting}

\begin{table}[t]
    \caption{The learners we evaluate in different IL scenarios in the main text.}
    \vspace{-0.5em}
    \begin{center}
    \resizebox{1.0\linewidth}{!}{
    \begin{tabular}{llccc}
    \toprule
    \textbf{Category} & \textbf{Learner} & \textbf{Task} & \textbf{Domain} & \textbf{Class} \\
    \midrule
    Architecture & XdG \cite{masse2018alleviating} & \checkmark &  & \\
    \midrule
    \multirow{4}{*}{Regularization} & EWC \cite{kirkpatrick2017overcoming} & \checkmark & \checkmark & \\ & Online EWC \cite{schwarz2018progress} &\checkmark &\checkmark & \\ & SI \cite{zenke2017continual} &\checkmark &\checkmark & \\ & LwF \cite{li2017learning} &\checkmark &\checkmark & \\
    \midrule
    \multirow{6}{*}{Replay} & DGR \cite{shin2017continual} &\checkmark &\checkmark &\checkmark \\ & DGR with distillation \cite{van2019three} &\checkmark &\checkmark &\checkmark \\ & RtF \cite{van2018generative} &\checkmark &\checkmark &\checkmark \\ & ER \cite{rolnick2019experience} &\checkmark &\checkmark &\checkmark \\ & A-GEM \cite{chaudhry2018efficient} &\checkmark &\checkmark &\checkmark \\ & iCaRL \cite{rebuffi2017icarl} & & &\checkmark\\
    \bottomrule
    \end{tabular}}
    \end{center}
    \label{tab:evaluate_learners}
\end{table}

\textbf{Threat model}. Threat modeling is a critical process while designing adversarial attacks and defense\cite{carlini2019evaluating}. Specifically, a threat model contains a set of assumptions about the adversary's goals, knowledge, and capabilities, through which we can identify and evaluate potential threats of a system and more importantly, put forward some mitigation strategies to make it more secure. 

In this work, the adversary's goal we consider is to achieve untargeted backdoor attack. It means that samples from any task containing the trigger pattern will lead to misclassification without a specified target class, resulting in a significant drop of model performance on backdoor samples. It is also essential to ensure that the performance on benign samples are close to that in no-attack scenario for undetectability. For knowledge and capabilities, we assume that the adversary only knows the training set of the task currently being trained by the victim learner, leaving other previous tasks and subsequent tasks invisible. We consider poisoning-based backdoor attack, in which the only part an adversary can manipulate is the dataset, and just a small number of poisoned samples (\textit{i.e.} no more than $5\%$) are allowed to be injected to the training set. Additionally, the adversary cannot interfere with the training or inference process and model architectures.

\textbf{Experimental setting.} As shown in \cref{tab:evaluate_learners}, we evaluate 11 typical incremental learners, including three categories which are based on architectural adjustment, regularization and replay, respectively. Following the definition in \cite{van2019three}, for each learner, we evaluate its robustness in the following three scenarios:
\begin{itemize}
    \item Task-IL: multiple tasks are learnt sequentially and the task identities of test data are always given.
    \item Domain-IL: training data of one task arrive in batches and each batch contains samples from all classes.
    \item Class-IL: classes in one task are learnt sequentially.
\end{itemize}

In this section, our experiments are based on permuted MNIST dataset, in which a new task is produced by shuffling pixels of samples in the MNIST dataset according to a random permutation with a total of 10 tasks. \textit{Evaluations on CIFAR-10 and CIFAR-100 datasets can be found in the appendix}. We adopt the typical poisoning-based backdoor attack framework named BadNets\cite{gu2019badnets}, and a $6\times 6$ checkerboard trigger is overlaid to the upper left corner of clean samples, as shown in \cref{fig:introduce} (right). This trigger can be further split in our distributed attack experiments. 

To investigate the spread of adverse effect across tasks, we implement data poisoning at the first and the last task respectively, and the index of its first class is set as the label of poisoned samples. Finally, we record the performance of the infected model on benign and malignant test samples from all tasks for comparison. Due to the weaker constraint of regularization-based learners compared with replay-based learners, they perform poorly in the hardest Class-IL scenario\cite{van2019three}, and thus are not considered in the corresponding experiments. Specifically, the IL learners we evaluate in different learning scenarios are listed in \cref{tab:evaluate_learners}. Unless otherwise specified, we keep the same hyper-parameter settings as in \cite{van2019three}. All the experiments below are repeated three times, and the random seeds are set to 0, 6666 and 8888, respectively.\footnote{We apply the implementation of each incremental learner in this repository: \href{https://github.com/GMvandeVen/continual-learning}{https://github.com/GMvandeVen/continual-learning}.}

Next, we show that incremental learners are vulnerable to backdoor attack, and the adverse effect can be easily spread forward and backward across tasks (\textit{i.e.}, the backdoor can be successfully triggered on any tasks), even if the proportion of poisoned samples is extremely small (e.g., $0.1\%$) or in the Task-IL scenario where different tasks are better isolated, expecting for an effective remedy.

\subsection{On the susceptibility of incremental learners}
\label{sec:susceptibility}

\begin{figure*}[t]
    \centering
    \includegraphics[width=0.95\linewidth,trim=0 0 0 0]{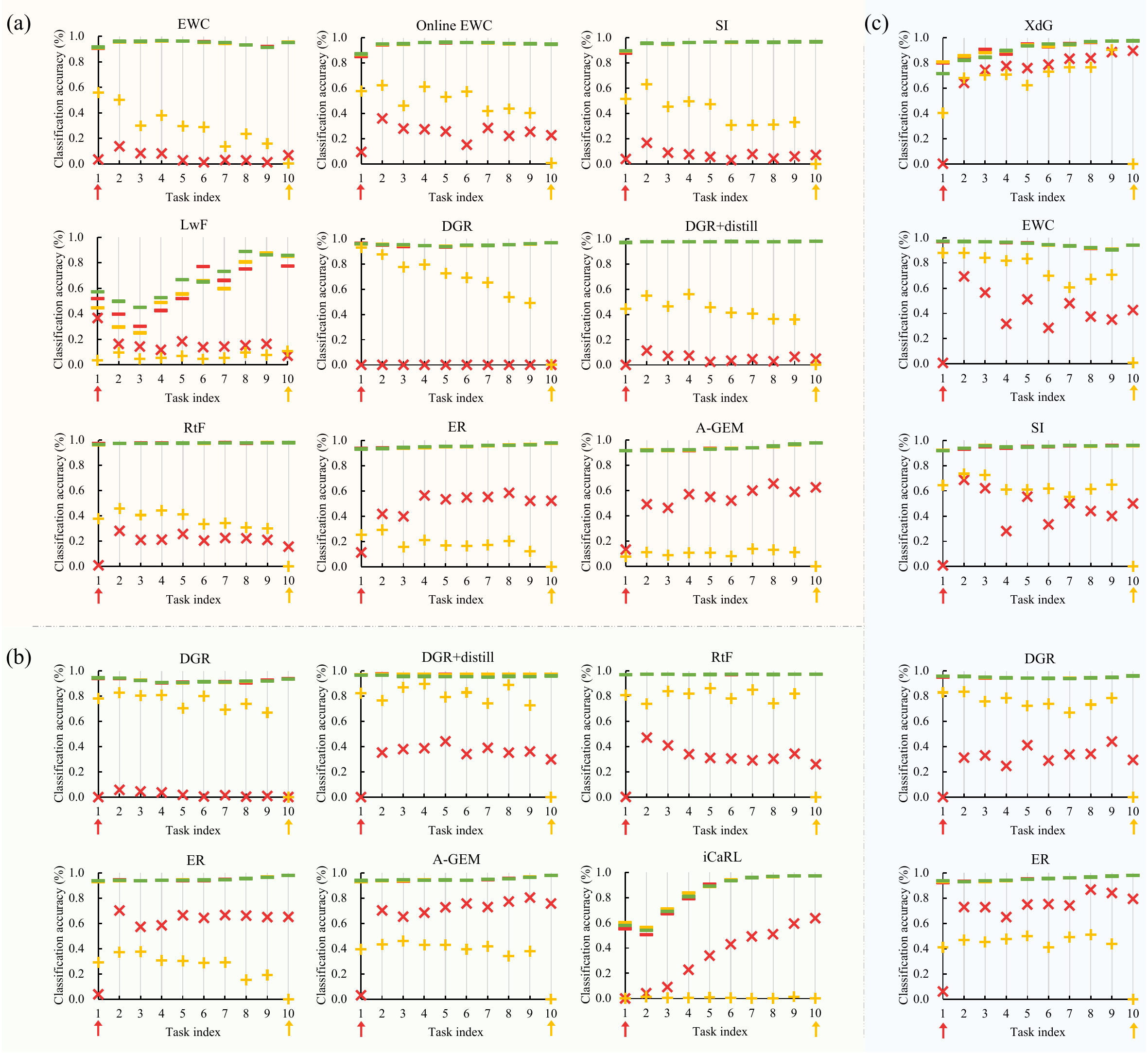}
    \caption{Performance of each IL learner on normal [$-$] and backdoor samples [$\times|+$] \textcolor[RGB]{112,173,71}{without attack} or after data poisoning [$\uparrow$] at the \textcolor{red}{beginning} or the \textcolor[RGB]{255,192,0}{ending} of Domain-IL (a), Class-IL (b) and Task-IL (c) task protocols on permuted MNIST dataset.}
    \label{fig:domain_class}
\end{figure*}

We first investigate the robustness of each incremental learner under Domain-IL and Class-IL task protocols only, as the Task-IL scenario is sometimes impractical and the task identities provided may result in better isolation between tasks which makes the cross-task generalization of adverse effect harder. 

We poison $5\%$ of the training set. Specifically, we randomly select $5\%$ of the training samples from the current task being trained, overlay the upper left corner of each sample with the trigger pattern displayed in \cref{fig:introduce}, change their labels to the first class of the current task, and finally add them to the original training set. As the training process completes, we report the performance of each learner on all tasks given clean or backdoor samples, and we also show the performance of models trained without attack for comparison. Our experimental results are shown in \cref{fig:domain_class}.

As we expected, for all incremental learners, we can successfully destroy the performance of the current task and achieve results which are comparable to conventional data poisoning scenarios---accuracy drops to nearly zero. In addition, we observe that the adversary can indeed affect other tasks and the backdoor can be successfully triggered, even if they are unknown while conducting data poisoning, indicating that the impact of our backdoor attack has the cross task propagating effect. A rough understanding for this feature is that, on one hand, the goal of incremental learning is to keep the knowledge of old tasks from being forgotten while learning new tasks, including backdoor knowledge, which result in the long-term impact of data poisoning. On the other hand, some strategies adopted by learners, such as regularization and replay, break the isolation nature between different tasks during the training process, directly or indirectly, which further promotes the diffusion of adverse effect produced by poisoned samples. 

\begin{figure}[t]
    \centering
    \includegraphics[width=0.75\linewidth,trim=0 0 0 0]{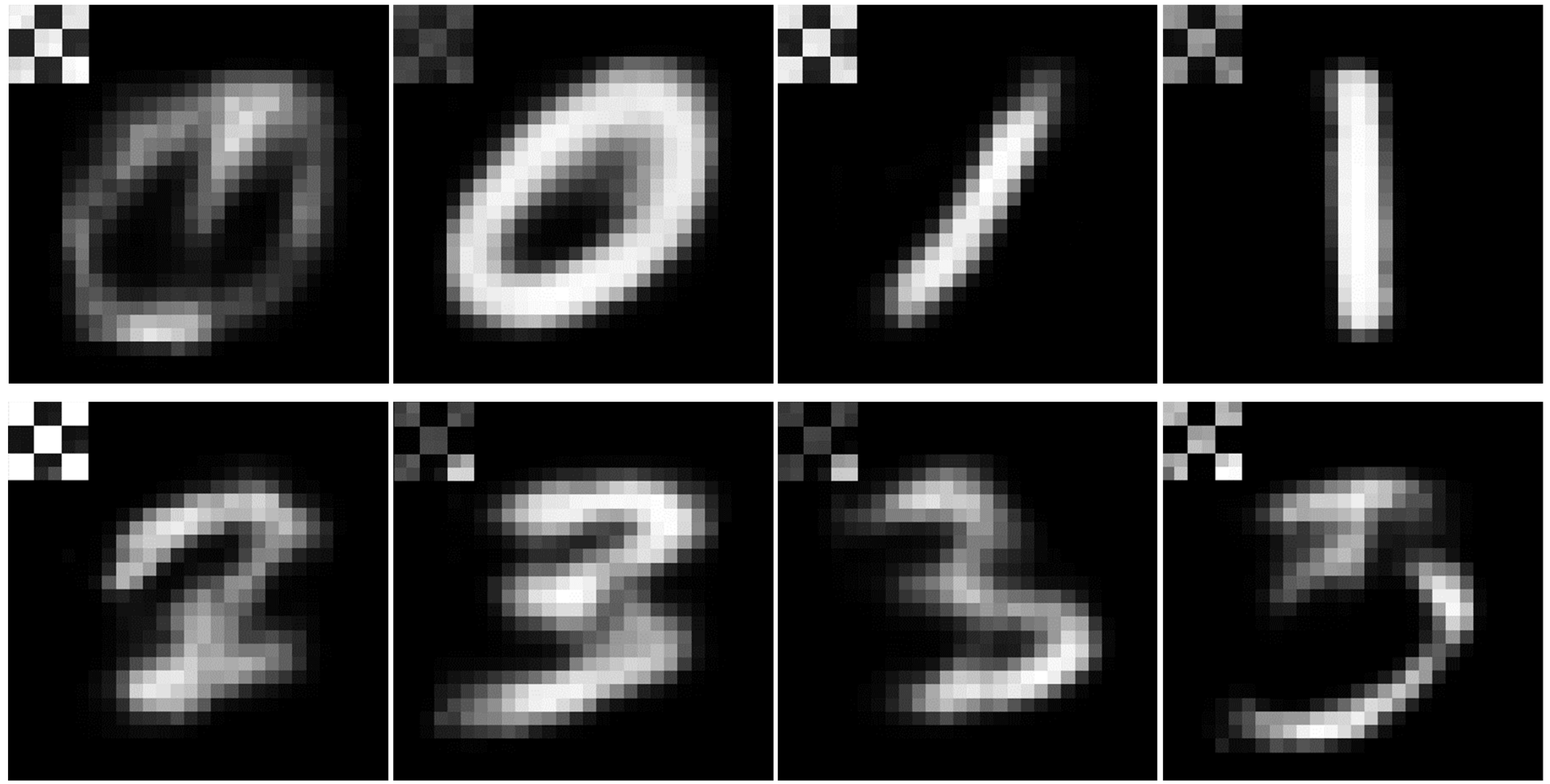}
    \caption{The replayed samples produced by the generator in DGR learner during the training process of the $3^{\text{rd}}$ task after we conduct data poisoning at the $1^{\text{st}}$ task of the split MNIST task protocol which contains only digits `0' and `1'.} 
    \label{fig:dgr_example}
\end{figure}

For instance, DGR \cite{shin2017continual} is a replay-based learner, and the samples to be replayed are generated by a variational autoencoder (VAE) which is trained along with the main model to fit the mixed data distribution of previous tasks. When a new task being trained, the model is jointly optimized on new samples and replayed samples, which breaks the barriers between tasks. Therefore, the adversary can successfully affect all existing tasks just like what happens in the conventional learning scenario where the samples from all classes are utilized altogether. Meanwhile, the VAE can also be misguided by the poisoned data, and thus the trigger pattern will consistently exist in replayed samples during the training process of subsequent tasks. We verified this feature under split MNIST task protocol\cite{van2019three}. Interestingly, as shown in \cref{fig:dgr_example}, in addition to the task we selected to conduct data poisoning (\textit{e.g.}, the first task containing digits `0' and `1'), the replayed samples generated by VAE for other tasks (\textit{e.g.}, digits `2' and `3' from the second task) also contain the trigger pattern. 

To better understand the propagation effect, we further noticed that, while adverse effect spread over tasks, backward propagation seems to encounter greater resistance than forward propagation. Take DGR as an example, as shown in \cref{fig:domain_class}, if we conduct data poisoning during the training phase of the first task, performance of all the subsequent tasks on backdoor samples drop to nearly zero. Meanwhile, however, if we instead do the same for the last task, performance of its previous tasks only drop by about $30\%$. 

The above phenomenon may stem from the influence of incremental learners on model stability and plasticity, which is the underlying cause of the susceptibility of incremental learners. Specifically, after we add the trigger on a clean image, the model's output depends on its main focus of attention, which is the result of the game between backdoor and benign knowledge. Before data poisoning, the model only relies on benign knowledge, and according to the stability requirement of incremental learning, this part of the benign knowledge can be sustained during the training process of subsequent tasks even if trigger patterns appear, and finally be the dominant component during the inference phase. This feature enables benign knowledge to compete with backdoor knowledge, which ends up with the difficulty of backward propagation. On the contrary, due to the plasticity requirement and shortcut tendency of DNNs\cite{geirhos2020shortcut}, if a new task being trained come under attack, the trigger pattern will be fitted by the model immediately and shows a significantly higher priority than benign features, which is the same as what happens when we conduct data poisoning in the conventional learning scenario. And during the training phase of subsequent tasks, the model's attention will continue to be dominated by features related to the trigger pattern under the constraint of stability requirement and the strongly misleading fact of triggers. Although the backward propagation seems harder, we argue that the adversary do not need to worry about it in the cyclic institutional incremental learning scenario where any previous institution can also be considered as a subsequent institution.

\subsection{Even the Task-IL scenario is sometimes brittle}

During the inference phase of the Task-IL scenario, the task identities are always provided. Therefore, this is the least practical scenario, since different tasks can be considered separately, such as using different set of parameters or even different models for different tasks to solve the catastrophic forgetting issue completely. For instance, the XdG learner\cite{masse2018alleviating} achieves isolation of parameters by using a separate sub-network for each task. For other learners, the isolating effect can be realized by leveraging a multi-head output layer as described in \cite{van2019three}. Therefore, it seems harder for different tasks to interact with each other, making it difficult for adverse effect to spread over tasks. 

However, as analyzed in \cref{sec:susceptibility}, another important factor that contributes to the backdoor-related transitivity is whether knowledge or samples from previous tasks present during the training process of subsequent tasks directly or indirectly, especially those related to the trigger pattern, which is a common characteristic of IL learners that are not based on parameter isolation, such as EWC and DGR. Meanwhile, not all learners for Task-IL scenario achieve complete isolation for different tasks, and their corresponding sub-networks partly overlap, which also expands the influence of backdoor knowledge.

Therefore, we follow the same setting as described in \cref{sec:setting} and redo the above experiments in the Task-IL scenario. Except for XdG, most of the learners listed in \cref{tab:evaluate_learners} are not specifically designed for this particular scenario. In order to take advantage of the given task identities, a feasible way is to expand the original model by a multi-head output layer where each task has its own independent output unit, so as to achieve the effect produced by parameter isolation\cite{van2019three}. The performance of 5 representative victim learners are shown in \cref{fig:domain_class}. According to our results, it can be seen that except for the XdG learner with relatively smaller intersections of sub-networks for different tasks, the performance of other learners under attack is significantly degraded, and the spread of adverse effect is also shown, which proves that we need to pay attention to the security risk of backdoor attacks in all of the three scenarios in IL.

\subsection{Tiny amounts of poisoned samples is harmful}

\begin{figure*}[t]
    \centering
    \includegraphics[width=1.0\linewidth,trim=0 0 0 0]{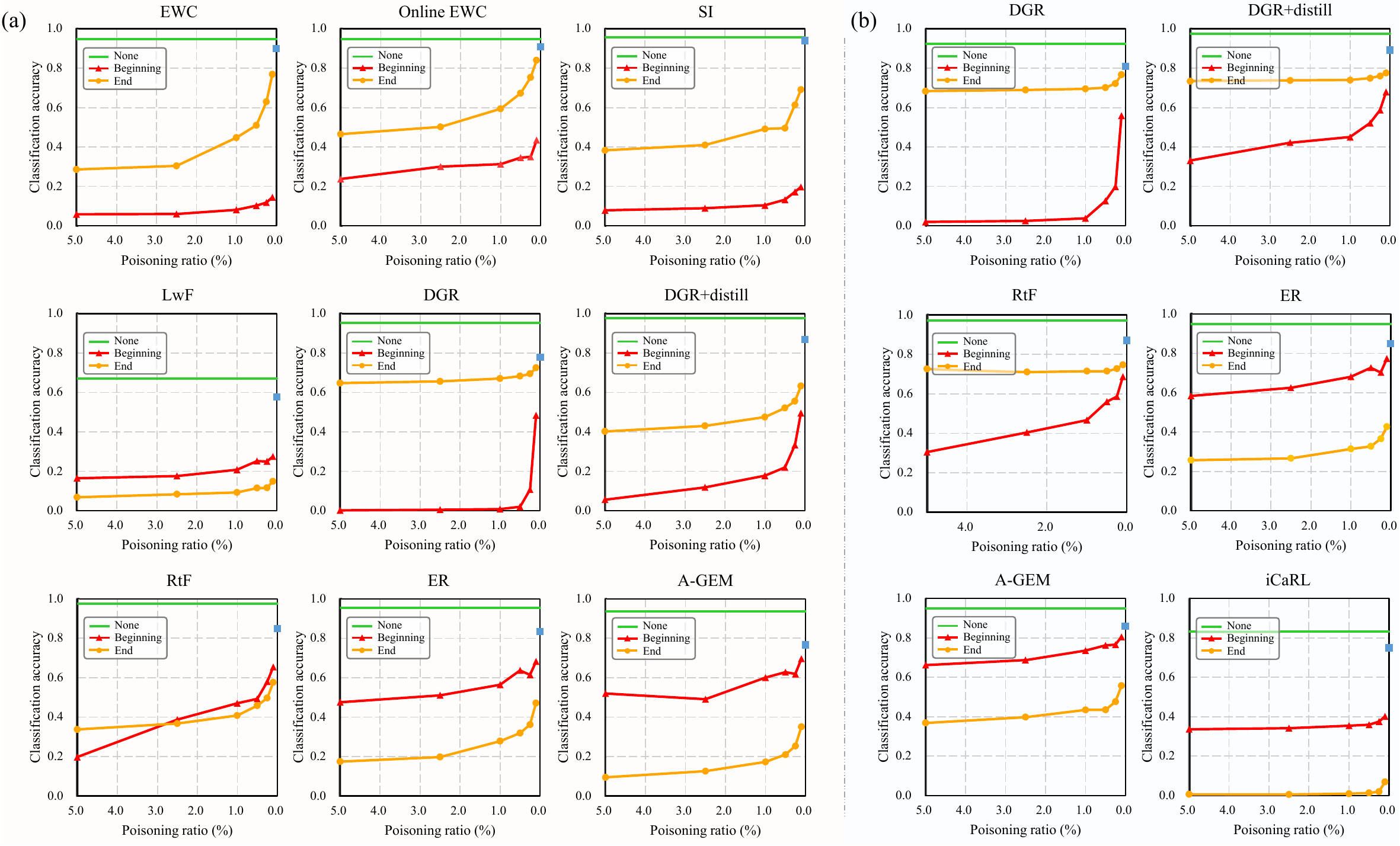}
    \caption{The average performance of each incremental learner on 10 tasks of Domain-IL (a) and Class-IL (b) task protocols on permuted MNIST dataset. We show the performance on \textcolor[RGB]{112,173,71}{clean} samples before attack and backdoor samples after data poisoning at the \textcolor{red}{beginning} or the \textcolor[RGB]{255,192,0}{ending} of each task protocol while the poison ratio is set to $5\%$, $2.5\%$, $1\%$, $0.5\%$, $0.25\%$ and $0.1\%$, respectively. We also show the performance on backdoor samples without data poisoning during the training process (\textit{i.e.}, the poison ratio is equal to $0\%$) for ablation study, which are marked as blue squares in the figure.}
    \label{fig:domain_class_ratio}
\end{figure*}

In this subsection, we impose stricter limits on the number of poisoned samples and observe whether the performance of the learners will still be degraded significantly. Specifically, we repeat the experiments in \cref{sec:susceptibility} by setting the poison ratio as $5\%$, $2.5\%$, $1\%$, $0.5\%$, $0.25\%$ and $0.1\%$, respectively. All experiments are repeated three times and the experimental results are reported in \cref{fig:domain_class_ratio}. Due to space limits, we only show the average accuracy of models on all tasks for each poison ratio. 

As can be seen, for most IL learners, we can consistently achieve satisfactory results when the poison ratio is greater than 1$\%$, and there is still a obvious drop in performance of each learner even if the ratio is set as low as $0.1\%$. For instance, in \cref{fig:domain_class_ratio} (a), after we conduct data poisoning during the training phase of task 1, the average accuracy of EWC learner on all tasks is only $14.48\%$, which is about $80\%$ lower than the baseline without attack. This observation is in line with what in conventional backdoor attack scenarios, demonstrating the significant misleading effect of trigger patterns, which is the result of shortcut learning characteristic of DNNs. Moreover, as will be analyzed in the next subsection, the adversary can also conduct distributed backdoor attacks to further reduce the poison ratio and increase the damaging effect.

\subsection{Distributed backdoor attack}

\begin{figure}[t]
    \vspace{0.5em}
    \centering
    \includegraphics[width=0.95\linewidth,trim=0 0 0 0]{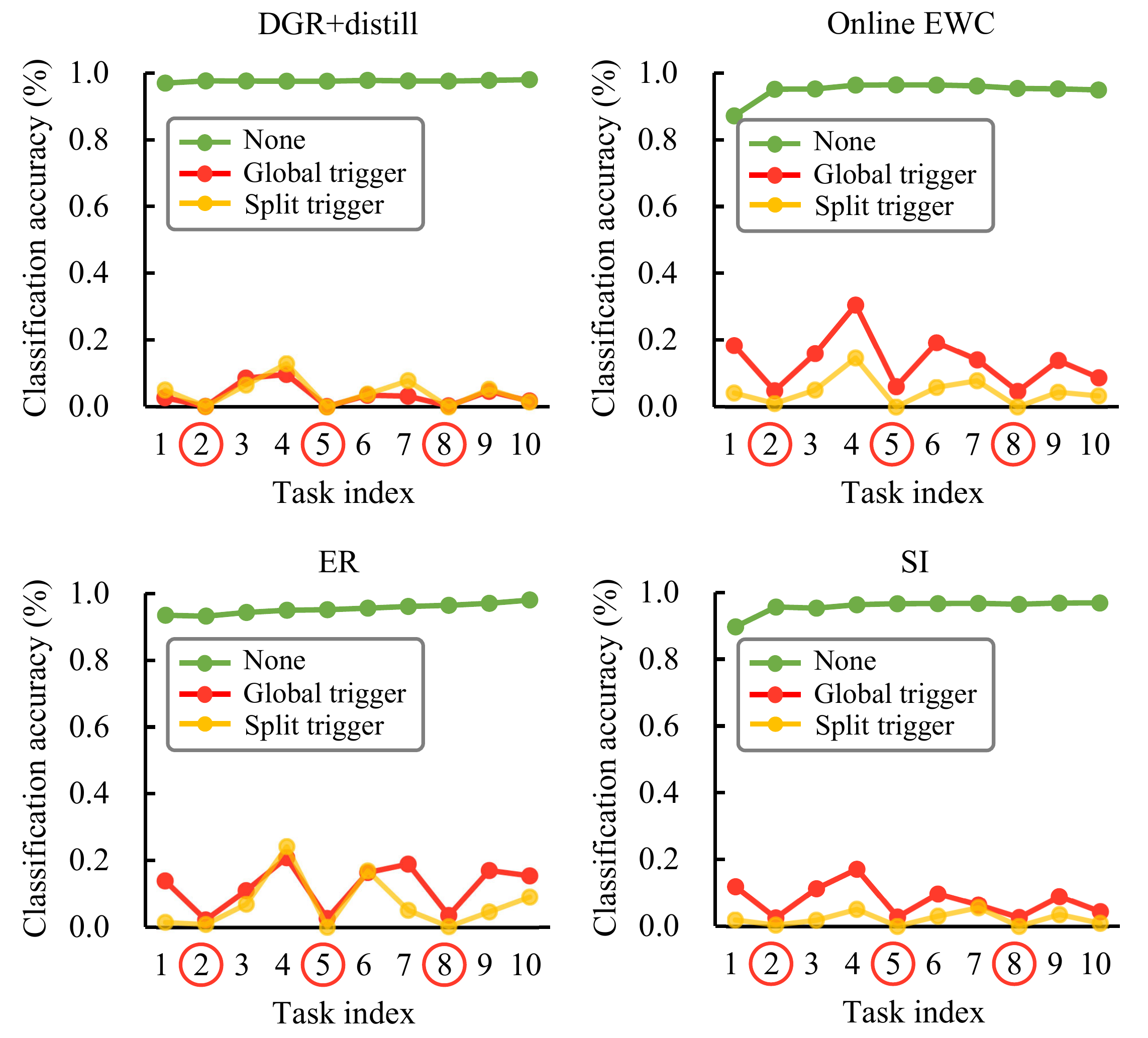}
    \vspace{-0.3em}
    \caption{The performance of 4 learners on clean and backdoor samples after data poisoning at tasks 2, 5 and 8 of Domain-IL task protocol on permuted MNIST dataset using the global trigger and split triggers respectively.}
    \label{fig:multi_task_poison}
\end{figure}

In this subsection, we show that the nature of streaming data in IL and the demonstrated spreading effect provide great convenience to the adversary. As opposed to conventional settings, in the field of IL, data arrives in batches over a time series and each batch may be poisoned by the adversary, which greatly increases the possibility of being attacked. Suppose that the number of tasks being attacked is a random variable $X$ which follows the binomial distribution $B(n,p)$ where $n$ is the number of tasks to be learned and $p$ is the probability of each task being successfully poisoned, then the probability that the learner is compromised (at least one task get infected) is $1-(1-p)^n$, and the expectation of $X$ is $np$, both of which increase with the number of tasks $n$, thus increasing the security risk of learners. What's worse, if an adversary is able to poison multiple tasks, the poison ratio can be set even lower, such as 1$\%$, while resulting in a devastating effect on the victim learner. 

We experimentally verified the feasibility of this distributed backdoor attack. We set $p$ to 0.3, and $n$ is 10 in the permuted MNIST task protocol. According to the above analysis, the adversary have a 97.17$\%$ chance to break the learner compared to 30$\%$ in conventional learning scenarios, and there are 3 tasks on average which can be compromised. We assume that the identities of the compromised tasks are 2, 5 and 8, respectively. The poison ratio is set to $1\%$. To comply with the threat model described in \cref{sec:setting}, the labels of the poisoned samples for each task are set to its first class. As can be seen from the model performance shown in \cref{fig:multi_task_poison} (red line), we achieve comparable or even better results compared to \cref{fig:domain_class} by leveraging $80\%$ fewer poisoned samples in each task. 

Another benefit of distributed attacks is that the trigger patterns in poisoned samples can be made stealthier. As described in \cite{xie2019dba}, the global trigger can be divided into multiple sub-patterns, and one of them is used for each compromised node. Finally, the original global trigger used during the inference phase can achieve comparable results to the non-split setting. We find it also practical in our scenario. Specifically, for each compromised task, the injected trigger is a sub-pattern of the global checkerboard with shape 2$\times$6, as illustrated in the poisoned samples for institution 1 and 3 in \cref{fig:introduce}. We repeat the above experiment with split triggers and the same hyper-parameters. The results shown in \cref{fig:multi_task_poison} (yellow line) demonstrate the effectiveness of distributed trigger patterns for IL.

\subsection{Evaluation Results on CIFAR datasets}

Based on our extensive experiments on permuted MNIST dataset, we conclude that in addition to the shortcut tendency of DNNs, the susceptibility of IL also come from the cross-task generalization effect of backdoor knowledge and more opportunities for data poisoning given to the adversary caused by the nature of streaming data, which is fundamentally different from conventional backdoor attack scenarios. For better completeness of the experiment, we further test the cross-task generalization of adverse effect and the feasibility of distributed attack on split CIFAR-10 and CIFAR-100 dataset which are also commonly used in IL research.

Our experimental results on CIFAR datasets are shown in the appendix. It is clear that the features we found on the split MNIST dataset still exist. However, the attack strength seems to become weaker. One possible explanation is that, there are two prerequisites for the success of backdoor attacks against IL. On one hand, the shortcut tendency of DNNs ensures that the poisoned tasks can be successfully broken. On the other hand, one of the goals of IL is to keep the old knowledge from being forgotten while learning new tasks, including backdoor knowledge, which result in the long-term impact of data poisoning. This conclusion is widely verified in the MNIST-based experiments because the existing learners perform very well. However, as can be seen from the experimental results on the split CIFAR-10/100 dataset, these tasks are far from been solved. The learners are not capable of handling such complex tasks, which makes it difficult to achieve the stability goal of IL. We suspect that the backdoor related knowledge is also affected by severe catastrophic forgetting issue, which is the main reason why the cross-task generalization of adverse effect becomes weaker on these datasets. Therefore, we believe that more advanced IL learners on CIFAR datasets in the future will still show obvious vulnerabilities against backdoor attack.

\section{How to detect the backdoor attack?}

\begin{figure}[t]
    \centering
    \vspace{0.5em}
    \includegraphics[width=0.95\linewidth,trim=0 0 0 0]{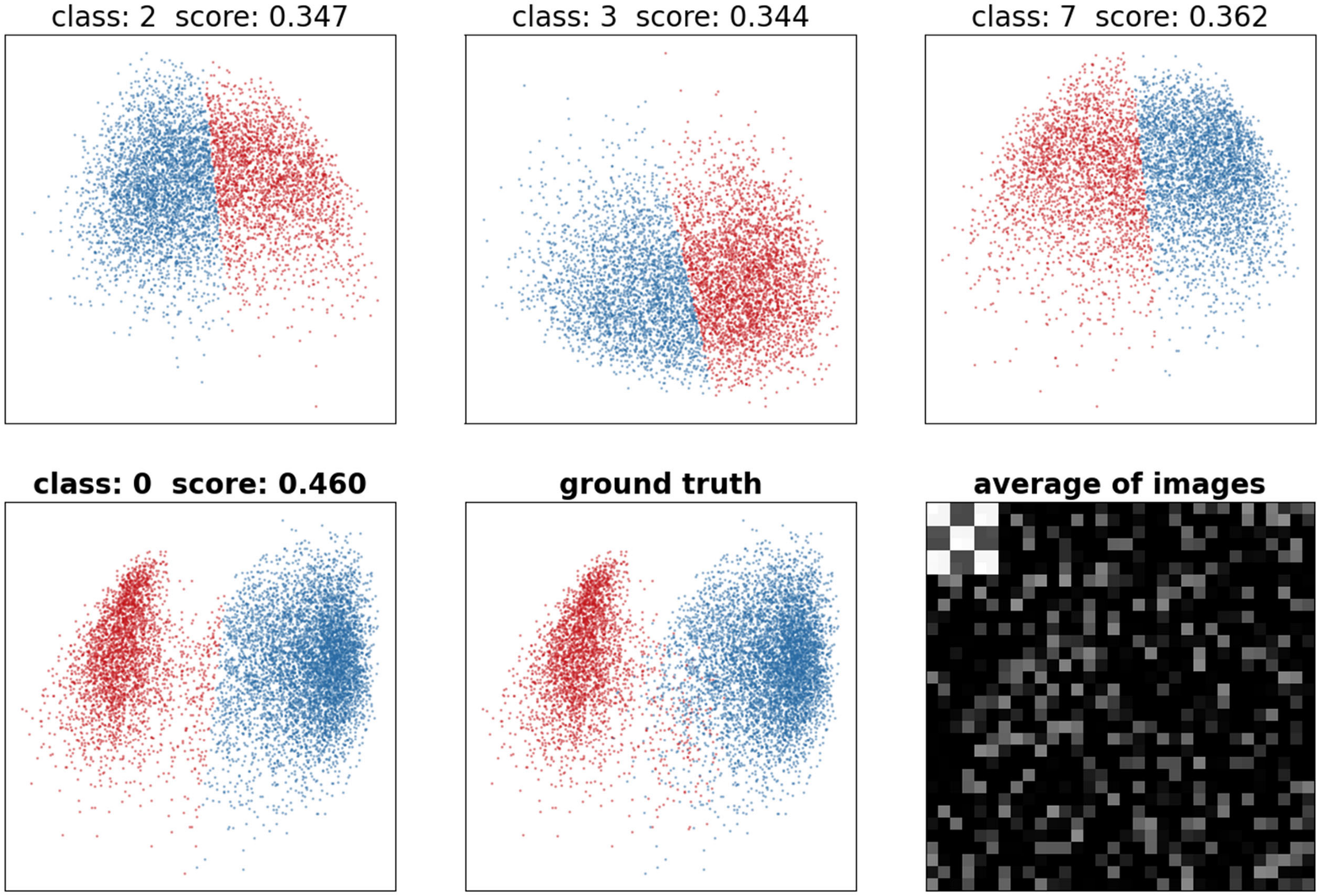}
    \caption{The results of $2$-means clustering for reduced activations of the penultimate hidden layer. The predicted class with the highest silhouette score as well as the average of images corresponding to one of its clusters reveal the label and the trigger pattern used by the adversary.} 
    \label{fig:defense}
\end{figure}

In the above subsections, we demonstrate that our flexible and damaging distributed backdoor attack is feasible for IL, which poses a security threat to applications bases on it, especially those with high security requirements, such as collaborative medical diagnoses\cite{chang2018distributed}. In view of this, an effective defense strategy should be provided. Backdoor attacks are usually inconspicuous, because no one knows the trigger patterns except the adversary, and the infected models behave normally on clean samples. Therefore, we need to detect triggers actively to fix our model.

Activation clustering\cite{chen2018detecting} is the first universal and effective method to detect poisoned samples without verified and trusted data, which is applicable for our scenario. The basic principle of activation clustering is that, even if the model predicts the same for clean samples and backdoor samples, the model's focus of attention, which can be reflected by the neural network activations, must be different. To take advantage of this feature, the authors find out malignant samples by leveraging cluster analysis on the activations of the last hidden layer .

In our scenario, we follow the procedure described in \cite{chen2018detecting} to detect backdoor attack after the training process of each task. Specifically, we first divide the training set into subsets according to the predictions of the model. For the samples in each subset, we obtain their corresponding set of network activations of the penultimate hidden layer (we find it more practical in our scenario). In order to facilitate the subsequent clustering, these activations are reduced to two-dimensional vectors by Independent Component Analysis (ICA). Finally, we perform cluster analysis on these reduced activations by $k$-means algorithm, where $k=2$. In our attack, all the poisoned samples for each task are mislabeled as the same class, while at the same time, the model produces different activate patterns on the poisoned samples and clean samples from the target class. Therefore, if the quality of cluster analysis for a certain class is high, which can be indicated by a high silhouette score, and the number of samples in one of its clusters is approximately the product of the poison ratio and size of the training set, then we can conclude that this cluster contains poisoned samples, thereby the trigger pattern is found. Finally, we correct the labels of all found poisoned samples and use them to fine-tune the model. We repeat this process after finishing the training for each task to eliminate the adverse effect of backdoor attacks.

We empirically evaluate the effectiveness of activation clustering. We attack DGR learner under Domain-IL task protocol, and the poisoning is conducted during the training phase of the $10^{th}$ task. The poison ratio is set to $5\%$, \textit{i.e.}, we randomly select 3000 samples from the training set to introduce the trigger pattern and then set their labels to 0. From the results of cluster analysis after finishing the training of the $10^{th}$ task shown in \cref{fig:defense}, the predicted class with the highest silhouette score is exactly 0, and the average of images corresponding to its red cluster restores the trigger pattern precisely. Meanwhile, the divided clusters for class 0 is close to the ground truth. There are 3021 samples in the red cluster, and 2896 of which are truly poisoned, which implies that the recall is 96.53$\%$. Finally, the labels of the found poisoned samples can all be corrected to fix the infected model.

\section{Conclusion}

In this paper, we empirically evaluate the robustness of IL against backdoor attack. Our results suggest that IL brings more opportunities and approaches to the backdoor adversary, which poses serious security risks. Our findings inspire future research to further explore more security vulnerabilities about IL and their corresponding defenses.


{\small
\bibliographystyle{ieee_fullname}
\bibliography{IL_backdoor}
}

\newpage

\section*{Appendix}
\label{appendix}

\appendix

\section{Experimental setting}
In this material, we reveal the vulnerability of IL on split CIFAR-10 and CIFAR-100 dataset. Taking advantage of the spreading effect, an adversary can carry out harmful distributed backdoor attack. Therefore, we experimentally verify these two features on each dataset. The trigger pattern and the hyper-parameter settings including poison ratios are consistent with that in the main text. All the experiments are repeated three times, and the finally average performance are reported. We apply the implementation of each incremental learner in this repository: \url{https://github.com/GMvandeVen/continual-learning}. 

\section{Experiments on split CIFAR-10}
\label{append:cifar10}

\begin{figure}[t]
    \centering
    \includegraphics[width=0.85\linewidth,trim=0 0 0 0]{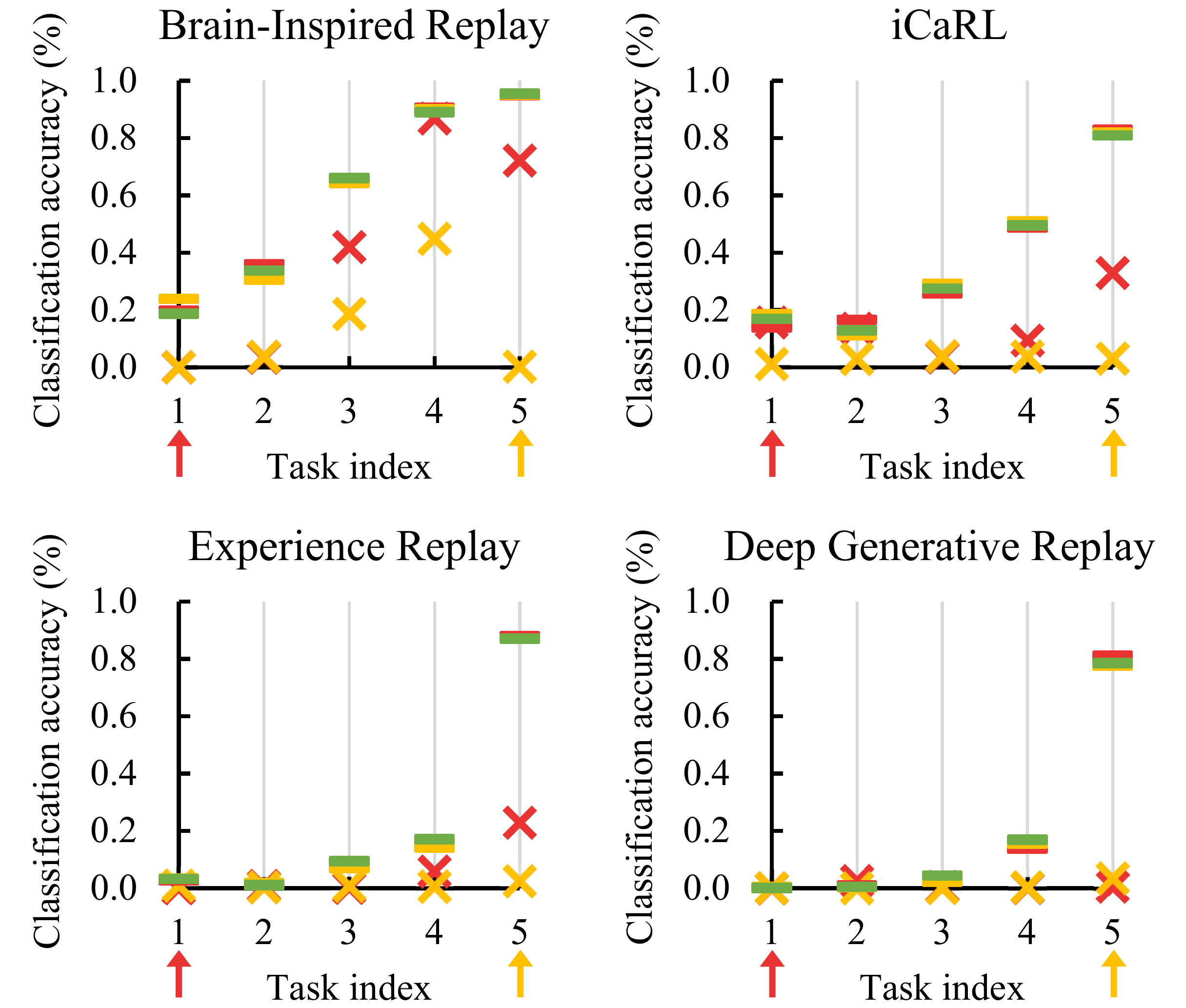}
    \caption{Performance of each learner on normal [$-$] and backdoor samples [$\times|+$] \textcolor[RGB]{112,173,71}{without attack} or after data poisoning [$\uparrow$] at the \textcolor{red}{beginning} or the \textcolor[RGB]{255,192,0}{ending} of Class-IL task protocol on split CIFAR-10 dataset.}
    \label{fig:cifar10}
\end{figure}

\begin{figure}[t]
    \centering
    \includegraphics[width=0.85\linewidth,trim=0 0 0 0]{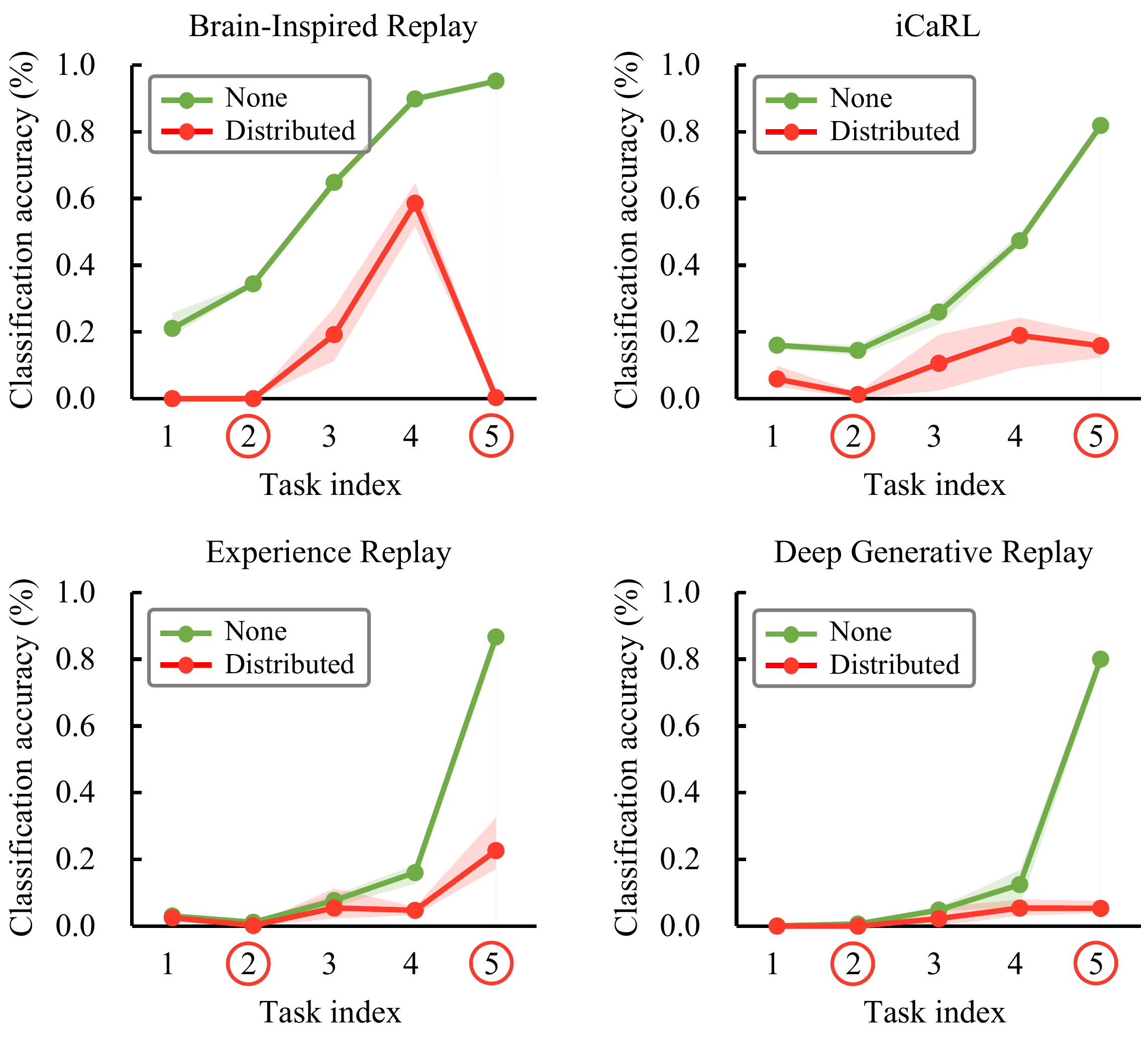}
    \caption{Performance of each learner on clean and backdoor samples after data poisoning at tasks 2 and 5 of Class-IL task protocol on split CIFAR-10 dataset using the split triggers.}
    \label{fig:cifar10_distributed}
\end{figure}

For split CIFAR-10, the original CIFAR-10 dataset is split into 5 tasks, and each task contains 2 classes, which is a binary classification problem. We evaluate our backdoor attack in the Class-IL scenario and select 4 learners that perform relatively well on this dataset for experiments, including Deep Generative Replay\cite{shin2017continual}, iCaRL\cite{rebuffi2017icarl}, Brain-Inspired Replay\cite{van2020brain} and Experience Replay\cite{rolnick2019experience}.

\textbf{Spread of adverse effect.}
We implement data poisoning at the $1^{st}$ and the $5^{th}$ task respectively. The label of poisoned samples are set to the first class of the selected task, \textit{i.e.}, 0 and 8. The results are reported in \cref{fig:cifar10}.

\textbf{Distributed attack.}
We assume that the identities of the compromised tasks are 2 and 5, and the corresponding label of poisoned samples are set to 2 and 8, respectively. We use the first 2 rows and the last 4 rows of the global trigger pattern as local triggers for task 2 and task 5, respectively. The results are reported in \cref{fig:cifar10_distributed}.

\section{Experiments on split CIFAR-100}

For split CIFAR-100, the original CIFAR-100 dataset is split into 10 tasks, and each task contains 10 classes. We select the Brain-Inspired Replay\cite{van2020brain} learner to evaluate our backdoor attack in the class-IL scenario.

\begin{figure}[t]
    \centering
    \includegraphics[width=0.95\linewidth,trim=0 0 0 0]{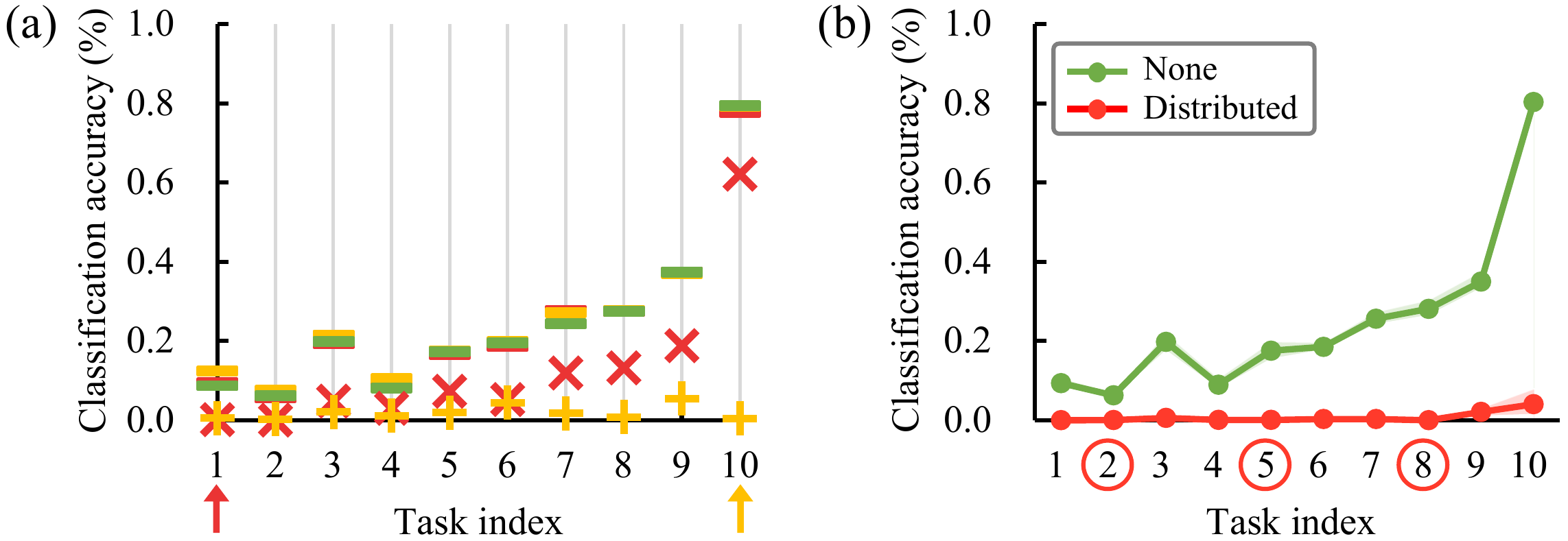}
    \caption{Results of the experiments on split CIFAR-100.}
    \label{fig:cifar100}
\end{figure}

\textbf{Spread of adverse effect.}
We implement data poisoning at the $1^{st}$ and the $10^{th}$ task respectively. The label of poisoned samples are set to the first class of the selected task, \textit{i.e.}, 0 and 90. The results are reported in \cref{fig:cifar100} (a).

\textbf{Distributed attack.}
We assume that the identities of the compromised tasks are 2, 5 and 8, and the corresponding label of poisoned samples are set to 10, 40, and 70, respectively. For each compromised task, we use two adjacent rows from the global trigger pattern as its local trigger. The results are reported in \cref{fig:cifar100} (b).

\end{document}